\documentclass[11pt]{article}
\usepackage{graphicx}
\usepackage{graphics}
\DeclareGraphicsExtensions{.eps}
\usepackage{float} 
\usepackage{amsmath}
\usepackage{amsfonts}   
\usepackage{amssymb}
\usepackage{subfloat}
\def\a{\alpha}

\begin{document}
\date{}
\title{{\bf{\Large Holographic $s$-wave condensate with non-linear electrodynamics: A nontrivial boundary value problem}}}

\author{
{\bf {\normalsize Rabin Banerjee}$
^{a}$\thanks{rabin@bose.res.in}},
{\bf {\normalsize Sunandan Gangopadhyay}$^{b,c}
$\thanks{sunandan.gangopadhyay@gmail.com, sunandan@bose.res.in}},
{\bf {\normalsize  Dibakar Roychowdhury}
$^{a}$\thanks{dibakar@bose.res.in, dibakarphys@gmail.com}},\, 
{\bf {\normalsize Arindam Lala}
$^{a}$\thanks{arindam.lala@bose.res.in, arindam.physics1@gmail.com}}\\
$^{a}$ {\normalsize  S.N. Bose National Centre for Basic Sciences,}\\{\normalsize JD Block, 
Sector III, Salt Lake, Kolkata 700098, India}\\
$^{b}$ {\normalsize Department of Physics, West Bengal State University, Barasat, India}\\
$^{c}${\normalsize Visiting Associate in Inter University Centre for Astronomy \& Astrophysics,}\\
{\normalsize Pune, India}
\\[0.3cm]
}
\date{}

\maketitle
\begin{abstract}
In this paper, considering the probe limit, we analytically study the onset of holographic $s$-wave condensate in the planar Schwarzschild-AdS background. Inspired by various low energy features of string theory, in the present work we replace the conventional Maxwell action by a (non-linear) Born-Infeld (BI) action which essentially corresponds to the higher derivative corrections of the gauge fields. Based on a variational method, which is commonly known as the Sturm-Liouville (SL) eigenvalue problem and considering a non-trivial asymptotic solution for the scalar field, we compute the critical temperature for the $s$-wave condensation. The results thus obtained analytically agree well with the numerical findings\cite{hs19}. As a next step, we extend our perturbative technique to compute the order parameter for the condensation. Interestingly our analytic results are found to be of the same order as the numerical values obtained earlier. 
\end{abstract}
\section{Introduction}
For the past few years, the AdS/CFT duality\cite{adscft1}-\cite{adscft2}, which provides an exact correspondence between a gravity theory in (d+1)-dimensional AdS space to that with a strongly coupled gauge theory living on d-dimensions has been extensively applied in order to describe various phenomena in usual condensed matter physics including high $T_{c}$ superconductivity. The holographic description of $s$-wave superconductors basically consists of a charged planar AdS black hole minimally coupled to a complex scalar field. The formation of scalar hair below the critical temperature ($T_{c}$) triggers the superconductivity in the boundary field theory through the mechanism of spontaneous $U(1)$ symmetry breaking\cite{hs1}-\cite{hs5}.

Besides the conventional framework of Maxwell electrodynamics, there is always a provision for incorporating non-linear electrodynamics in various aspects of gravity theories. The theory of non-linear electrodynamics was originally introduced in an attempt to remove certain discrepancies, such as the infinite self energy of electrons, for the Maxwell theory\cite{Born-Infeld}. Recently gravity theories with non-linear electrodynamics have found profound applications due to its emergence in the low energy limit of the heterotic string theory\cite{Kats}-\cite{Cai}. Gravity theories that include such non-linear effects have been investigated extensively for the past several years\cite{ABG1}-\cite{Lala}. As a result, several intriguing features regarding the properties of black holes, such as regular black hole solutions\cite{ABG1},\cite{ABG2}, validation of the zeroth and the first law of black hole mechanics\cite{Rasheed}, different asymptotic behaviours of black hole solutions\cite{Has4} etc., have been emerged. 

Among the various theories with non-linear electrodynamics, it is the Born-Infeld (BI) theory that has attained renewed attentions due to its several remarkable features. Perhaps the most interesting and elegant regime for the application of the BI electrodynamics is the string theory. The BI theory effectively describes the \textit{low energy} behaviour of the D-brane which are basically (non-perturbative) solitonic objects in string  theory\cite{D_1}-\cite{D_2}. The non-linear theories have entered into the gauge theories via the ``brane-world" scenario\cite{Gibbons1_JHEP}. String theory requires the inclusion of gravity theories in order to describe some of its fundamental properties. In this regard it is indeed essential to connect non-linear electrodynamics with gravity. One of the interesting properties of the BI theory is that the electric field is regular for a point-like particle. The regular BI theory with finite energy gives the non-singular solutions of the field equations. In fact the BI electrodynamics is the only non-linear electrodynamic theory with a sensible weak field limit\cite{Boillat, Gibbons2}. Another intriguing feature of the BI theory is that it remains invariant under electromagnetic duality\cite{Gibbons1_JHEP}, \cite{Hatsuda}-\cite{Khare}. All the above mentioned features of the BI theory provide a motivation to study Einstein gravity as well as higher curvature gravity theories coupled to BI electrodynamics\cite{ABG2},\cite{BI1}-\cite{BI8}. BI electrodynamics coupled to anti de-Sitter (AdS) gravity exhibits close resemblance to the Reissner-Nordstr\"om-AdS blck holes\cite{BI_RN1}-\cite{BI_RN2}. Also, it is reassuring to note that in $(3+1)$-dimensions the black hole solution with BI electrodynamics possesses a (lower) bound to the extremality of the BI-AdS black holes\cite{Dibakar}-\cite{Lala},\cite{Cai00}.

For the past couple of years gravity theories with both linear (usual Maxwell case) as well as non-linear electrodynamics have been extensively studied in the context of AdS/CFT superconductivity\cite{hs6}-\cite{hs27}. Surprisingly it is observed that lesser efforts have been paid while dealing with non-linear theories\cite{hs19}-\cite{hs27}. The analysis that have been performed so far are mostly based on numerical techniques. A systematic analytic approach is therefore lacking in this particular context.

Inspired by all the above mentioned facts, in the present paper we aim to study the onset of holographic $s$-wave condensate in the framework of BI electrodynamics. In fact, this issue has been investigated earlier in \cite{hs19} using numerical techniques. In the present paper we aim to investigate the onset of $s$-wave condensate based on analytic technique which is popularly known as Sturm-Liouville (SL) eigenvalue problem\cite{hs7}. In the present analysis we adopt the boundary condition $\langle
\mathcal{O}\rangle=\psi^{+}$ and $\langle \mathcal{O}\rangle=\psi^{-}=0$
which implies that the conformal dimension of the condensation operator $\mathcal{O}$ in the boundary field theory is $\Delta_{+}=2$ \cite{hs19}. This boundary condition seems to be quite non-trivial as far as analytic computation is concerned. This is simply due to the fact that the perturbative technique required to solve the differential equations does not work in a straightforward manner. However, we overcome this problem by adopting certain mathematical tricks and have successfully computed the onset of $s$-wave condensate. From our analysis we have been (analytically) able to show that the critical temperature ($T_{c}$) indeed gets affected due to the presence of higher derivative corrections to the usual Maxwell action. In fact it is found to be decreasing with the increase in the value of the BI coupling parameter ($b$) which suggests the onset of harder condensation. It is also noteworthy that our analytic results are in good agreement with the existing numerical results\cite{hs19}. It should be noted that all our calculations have been carried out in the probe limit\cite{hs5,hs8}.\footnote{In the probe limit, gravity and matter decouple and the backreaction of the matter fields (the charged gauge field and the charged massive scalar field) is suppressed in the neutral black hole background. This is done by rescaling the matter fields by the charge ($q$) of the scalar field and then taking the limit $q\rightarrow\infty$. This simplifies the problem without hindering the physical properties of the system.}

The organization of the paper is as follows. In Section 2, the basic set up for holographic superconductors in the Schwarzchild-AdS background has been given. In Section 3 we have performed the analytic calculations involve to determine the critical temperature ($T_{c}$) of the condensate. Section 4 deals with the computation of the order parameter ($\langle
\mathcal{O}\rangle$) for the condensation. The last section is devoted to the conclusions. 

\section{Basic set up}
 To begin with, we consider a fixed planar Schwarzschild-AdS black hole background which reads\footnote{Without loss of generality we can choose $l=1$ which follows from the scaling properties of the equation of motion. Also, the gravitational constant is set to be unity ($G=1$).}\cite{hs6} 
\begin{equation}
ds^{2}=-f(r)dt^{2}+f(r)^{-1}dr^{2}+r^{2}(dx^{2}+dy^{2})
\label{e1}
\end{equation}

where the metric function is,
\begin{equation}
f(r)=\left(r^{2}-\frac{r_{+}^{3}}{r}\right),
\label{e2}
\end{equation}
$ r_{+} $ being the horizon radius of the black hole. 
The Hawking temperature of the black hole may be written as,
\begin{eqnarray}
T=\frac{1}{4\pi}\left(\frac{\partial f(r)}{\partial r}\right)_{r=r_{+}}=\frac{3r_{+}}{4\pi}.
\label{e3}
\end{eqnarray}

In the presence of electric field and a complex scalar field ($ \psi(r) $) we may write the corresponding Lagrangian density as
\begin{equation}
\mathcal{L}=\mathcal{L}_{BI}-|\nabla_{\mu}\psi -iqA_{\mu}\psi|^{2}-m^{2}|\psi|^{2}
\label{e4}
\end{equation}

where $ \mathcal{L}_{BI} $ is the Born-Infeld Lagrangian density given by\cite{hs19},
\begin{equation}
\mathcal{L}_{BI}=\frac{1}{b}\left(1-\sqrt{1+\frac{bF}{2}}\right).
\label{e5}
\end{equation}
Here $ F\equiv F^{\mu\nu}F_{\mu\nu} $ and $ b $ is the Born-Infeld parameter. It is to be noted that, in our approach we shall investigate the effect of the higher derivative corrections to the gauge field in the leading order i.e. we keep terms only linear in $b$. Thus, the results of this paper is valid only in the leading order of $b$.

The equation of motion for the electromagnetic field tensor $F_{\mu\nu}$ can be written as,
\begin{equation}
\partial_{\mu}\left(\frac{\sqrt{-g}F^{\mu\nu}}{\sqrt{1+\frac{bF}{2}}}\right)=\mathcal{J}^{\nu}.
\label{e6}
\end{equation}
Considering the ansatz\cite{hs6} $ \psi=\psi(r) $ and $ A_{\mu}=(\phi(r),0,0,0) $ the equations of motion for the scalar field $ \psi(r) $ and the electric scalar potential $ \phi(r) $ may be written as,\footnote{In our analysis we take $m^{2}=-2$.} 
\begin{equation}
\psi ''(r)+\left(\frac{f'}{f}+\frac{2}{r}\right)\psi'(r)+\left(\frac{\phi^{2}(r)}{f^{2}}+\frac{2}{f}\right) \psi(r)=0
\label{e7}
\end{equation}

\begin{equation}
\phi''(r)+\frac{2}{r}\left(1-b\phi'^{2}(r)\right)\phi'(r)-\frac{2\psi^{2}(r)}{f}\left(1-b\phi'^{2}(r)\right)^{\frac{3}{2}}=0. 
\label{e8} 
\end{equation}

The above set of equations (Eqns.\eqref{e7}, \eqref{e8}) are written in the radial coordinate $r$. In order to carry out an analytic computation we define a new variable $z=\dfrac{r_{+}}{r}$. In this new variable Eqn.\eqref{e7} and Eqn.\eqref{e8} become
\begin{equation}
z\psi''(z)-\dfrac{2+z^{3}}{1-z^{3}}\psi'(z)+\left[z\dfrac{\phi^{2}(z)}{r_{+}^{2}(1-z^{3})^{2}}+\dfrac{2}{z(1-z^{3})}\right]\psi(z)=0, 
\label{e9} 
\end{equation}

\begin{equation}
\phi''(z)+\dfrac{2bz^{3}}{r_{+}^{2}}\phi'^{3}(z)-\dfrac{2\psi^{2}(z)}{z^{2}(1-z^{3})}\left(1-\dfrac{bz^{4}}{r_{+}^{2}}\phi'^{2}(z)\right)^{\frac{3}{2}}\phi(z)=0 .
\label{e10} 
\end{equation}

Since the above equations (Eqns.\eqref{e9},\eqref{e10}) are second order differential equations, therefore in order to solve them we must know the corresponding boundary conditions. The regularity of $\phi$ and $\psi$ at the horizon requires $ \phi(z=1)=0 $ and $ \psi(z=1)=\dfrac{3}{2}\psi'(z=1)$. 

\noindent On the other hand, at the spatial infinity $\phi$ and $\psi$ can be approximated as 
\begin{eqnarray}
\phi(z)&\approx & \mu - \frac{\rho}{r}\nonumber \\
&=& \mu - \frac{\rho}{r_{+}}z
\label{e11}
\end{eqnarray}
and 
\begin{eqnarray}
\psi(z)&\approx & \frac{\psi^{(+)}}{r^{\Delta_{+}}}+ \frac{\psi^{(-)}}{r^{\Delta_{-}}}\nonumber\\
&=&\frac{\psi^{(+)}}{r_{+}^{\Delta_{+}}}z^{\Delta_{+}}+ \frac{\psi^{(-)}}{r_{+}^{\Delta_{-}}}z^{\Delta_{-}}
\label{e12}
\end{eqnarray}
where $ \Delta_{\pm}=\frac{3}{2}\pm \sqrt{\frac{9}{4}+m^{2}} $ is the conformal dimension of the \textit{condensation operator} $ \mathcal{O} $ in the boundary field theory and $ \mu $ and $ \rho $ are interpreted as the chemical potential and charge density of the dual field theory. Since we have considered $m^{2}<0$ (which is above the BF bound\cite{BF1, BF2}), we are left with the two different condensation operators of different dimensionality corresponding to the choice of quantization of the scalar field $\psi$ in the bulk. In the present context either $\psi^{(+)}$ or $\psi^{(-)}$ will act as a condensation operator while the other will act as a source. In the present work we choose $\psi^{(+)}=\langle\mathcal{O}\rangle$ and $\psi^{(-)}$ as its source. Since we want the condensation to take place in the absence of any source, we set $\psi^{(-)}=0$. At this point, it must be stressed that for the present choice of $\psi$ the analytic calculations of various entities near the critical point get notoriously difficult and special care should be taken in order to carry out a perturbative analysis. In the present work we focus to evade the above mentioned difficulties by adopting certain mathematical techniques. Our analysis indeed shows a good agreement with the numerical results existing in the literature\cite{hs19}.
\section{$s$-wave condensate with non trivial boundary condition}
 With the above set up in place, we now move on to investigate the relation between the critical temperature of condensation and the charge density. 

 At the critical temperature $T_{c}$ the scalar field $\psi$ vanishes, so Eqn.(\ref{e10}) becomes
\begin{equation}
\phi''(z)+\frac{2bz^{3}}{r_{+(c)}^{2}}\phi'^{3}(z)=0.
\label{e13}
\end{equation}

The solution for this equation in the interval $[z, 1]$ reads\cite{Sun2} 
\begin{equation}
\phi(z)=\lambda r_{+(c)}\xi(z)
\label{e14}
\end{equation}
where
\begin{equation}
\xi(z)=\int_{z}^{1}\dfrac{d\tilde{z}}{\sqrt{1+b\lambda^{2}\tilde{z}^{4}}}.
\label{sol1}
\end{equation}

We shall perform a perturbative expansion of $b\lambda^{2}$ in the r.h.s of Eqn.(\ref{sol1}) and retain only the terms that are linear in $b$ such that $b\lambda^{2}=b\lambda_{0}^{2}+O(b^{2})$, where $\lambda_{0}^{2}$ is the value of $\lambda^{2}$ for $b=0$. Now for our particular choice of $\psi^{(i)}$ $(i=+,-)$ we have $\lambda_{0}^{2}\approx 17.3$ \cite{hs7}. Recalling that the existing values of $b$ in the literature are $b=0.1,\;0.2,\;0.3$ \cite{hs19} we observe that $b\lambda_{0}^{2}>1$. Consequently the binomial expansion of the denominator in (\ref{sol1}) has to done carefully. The integration appearing in Eqn.(\ref{sol1}) is done for two ranges of values of $z$, one for $z\leq \Lambda<1$ while the other for $\Lambda\leq z\leq 1$, where $\Lambda$ is such that $b\lambda_{0}^{2}z^{4}|_{z=\Lambda}=1$. At this stage, it is to be noted that $b\lambda_{0}^{2}z^{4}<1$ for $z<\Lambda$, whereas, on the other hand  $b\lambda_{0}^{2}z^{4}>1$ for $z>\Lambda$. 

For the first case ($z\leq \Lambda<1$),
\begin{eqnarray}
\xi(z)=\xi_{1}(z)&=&\int_{z}^{\Lambda}\dfrac{d\tilde{z}}{\sqrt{1+b\lambda_{0}^{2}\tilde{z}^{4}}}
+\int_{\Lambda}^{1}\dfrac{d\tilde{z}}{\sqrt{1+b\lambda_{0}^{2}\tilde{z}^{4}}}\nonumber\\
&\approx &\int_{z}^{\Lambda}\left(1-\frac{b\lambda_{0}^{2}\tilde{z}^{4}}{2}\right)+\frac{1}{\sqrt{b}\lambda_{0}}\int_{\Lambda}^{1}\left(\dfrac{1}{\tilde{z}^{2}}-\dfrac{1}{2b\lambda_{0}^{2}\tilde{z}^{6}}\right) \nonumber\\
&= &\left[\dfrac{9}{5}\Lambda -z+\dfrac{z^{5}}{10\Lambda^{4}}-\Lambda^{2}+\dfrac{\Lambda^{6}}{10}\right].
\label{e15}
\end{eqnarray}

Similarly in the range $\Lambda\leq z\leq 1 $, we have
\begin{eqnarray}
\xi(z)=\xi_{2}(z)&=&\int_{z}^{1}\dfrac{d\tilde{z}}{\sqrt{1+b\lambda_{0}^{2}\tilde{z}^{4}}}\nonumber\\
&\approx &\frac{1}{\sqrt{b}\lambda_{0}}\int_{z}^{1}\left(\dfrac{1}{\tilde{z}^{2}}-\dfrac{1}{2b\lambda_{0}^{2}\tilde{z}^{6}}\right)\nonumber\\
&= &\dfrac{\Lambda^{2}}{z^{5}}\left[z^{4}(1-z)+\dfrac{\Lambda^{4}}{10}(z^{5}-1)\right].
\label{e17}
\end{eqnarray}

From Eqn.\eqref{e17} one may note that the boundary condition $\phi(1)=0$ is indeed satisfied ($\xi_{2}(1)=0$).

We may now express $\psi(z)$ near the boundary as 
\begin{equation}
\psi(z)=\dfrac{<\mathcal{O}>}{\sqrt{2}r^{2}_{+}}z^{2}\mathcal{F}(z)
\label{e18}
\end{equation}
with the condition $\mathcal{F}(0)=1$ and $\mathcal{F}'(0)=0$.

Using Eqn.\eqref{e18} we may write Eqn.(\ref{e9}) as, 
\begin{equation}
\mathcal{F}''(z)-\dfrac{(5z^{4}-2z)}{z^{2}(1-z^{3})}\mathcal{F}'(z)-\dfrac{4z^{3}}{z^{2}(1-z^{3})}\mathcal{F}(z)+\lambda^{2}\dfrac{\xi^{2}(z)}{(1-z^{3})^{2}}\mathcal{F}(z)=0.
\label{e19}
\end{equation}

This equation can be put in the Sturm-Liouville form as,
\begin{equation}
\left[p(z)\mathcal{F}'(z)\right]'+q(z)\mathcal{F}(z)+\lambda^{2} g(z)\mathcal{F}(z)=0 
\label{e20} 
\end{equation}
with the following identifications,
\begin{eqnarray}
p(z)&=&z^{2}(1-z^{3})\nonumber\\ q(z)&=&-4z^{3}\nonumber\\
g(z)&=&\dfrac{z^{2}}{(1-z^{3})}\xi^{2}(z)=\chi(z)\xi^{2}(z)
\label{e21}
\end{eqnarray}
where, $\chi(z)=\dfrac{z^{2}}{(1-z^{3})}. $

Using Eqn.(\ref{e21}), we may write the eigenvalue $\lambda^{2}$ as,
\begin{eqnarray}
\lambda^{2}&=&\dfrac{\int_{0}^{1}\left\lbrace p(z)[\mathcal{F}'(z)]^{2}-q(z)[\mathcal{F}(z)]^{2}\right\rbrace dz}{\int_{0}^{1}\left\lbrace g(z)[\mathcal{F}(z)]^{2}\right\rbrace dz}\nonumber\\
&=&\dfrac{\int_{0}^{1}\left\lbrace p(z)[\mathcal{F}'(z)]^{2}-q(z)[\mathcal{F}(z)]^{2}\right\rbrace dz}{\int_{0}^{\Lambda}\left\lbrace \chi(z)\xi_{1}^{2}(z)[\mathcal{F}(z)]^{2}\right\rbrace dz+\int^{1}_{\Lambda}\left\lbrace \chi(z)\xi_{2}^{2}(z)[\mathcal{F}(z)]^{2}\right\rbrace dz}.
\label{e22}
\end{eqnarray}

We now choose the trial function $\mathcal{F}(z)$ as\cite{hs7}\footnote{The function $\mathcal{F}(z)$ as well as $\xi(z)$ (Eqn.\eqref{sol1}) do not appaer in the numerical analysis since in the numerical method Eqns.\eqref{e9} and \eqref{e13} are solved directly.} 
\begin{equation}
\mathcal{F}(z)=1-\a z^{2}
\label{e23}
\end{equation}
which satisfies the conditions $\mathcal{F}(0)=1$ and $\mathcal{F}'(0)=0$. This form of the trial function is also compatible with the boundary behaviour of the scalar field $\psi(z)$ (Eqn.\eqref{e12}).

Let us now determine the eigenvalues for different values of the parameter $b$. 

For $b=0.1$, we obtain
\begin{equation}
\lambda^{2}=300.769+\dfrac{2.27395\a -5.19713}{0.0206043+(0.00265985\a -0.0119935)\a}
\label{e24}
\end{equation}
which has a minima for $\a \approx 0.653219$. Therefore from eq.(\ref{e22}) we obtain
\begin{equation}
\lambda^{2}\approx 33.8298
\label{e25}
\end{equation}

The value of $\lambda^{2}$ obtained from the perturbative calculation justifies our approximation for computing the integral in Eqn.\eqref{sol1} upto order $b$ and neglecting terms of order $b^{2}$ and higher since the term of order $b^{2}$ can be estimated to be smaller than the term of order $b$.

Using Eqn.\eqref{e25}, the critical temperature for condensation ($T_{c}$) in terms of the charge density ($\rho$) can be obtained as, 
\begin{equation}
T_{c}=\dfrac{3r_{+(c)}}{4\pi}=\gamma \sqrt{\rho}\approx 0.099\sqrt{\rho}.
\end{equation}
where $\gamma = \dfrac{3}{4\pi\sqrt{\lambda}}$ is the coefficient of $T_{c}$. The value thus obtained analytically is indeed in very good agreement with the numerical result: $T_{c}=0.10072\sqrt{\rho}$ \cite{hs19}. Similarly, for the other values of the Born-Infeld parameter ($b$), we obtain the corresponding perturbative values for the coefficients of $T_{c}$ which are presented in the Table 1 below.
\begin{table}[h]
\centering                          
\begin{tabular}{c c c c}            
\hline\hline                        
Values of $b$ & $\gamma_{numerical}$ & $\gamma_{SL}$  \\ [0.05ex]
\hline
0.1 & 0.10072 &  0.099 \\ \\
0.2 & 0.08566 &  0.093  \\ \\
0.3 & 0.07292 &  0.089 \\  [0.05ex]           
\hline                                
\end{tabular}\label{T1}  
\caption{A comparison between analytic and numerical values for the coefficient ($\gamma$) of $T_{c}$ corresponding to different values of $b$}
\end{table}

Before concluding this section, we would like to emphasize the subtlety of the analytic method adapted here. We employ a perturbative technique to compute the integral in Eqn.\eqref{sol1} upto order $b$. This approximation is valid since we have investigated the effect of the higher derivative corrections upto the leading order in the nonlinear parameter ($b$). However, due to the nature of the integrand of Eqn.\eqref{sol1}, we had to be careful in separating the integral in two regions in order to perform a binomial expansion of the integrand.
\section{Order parameter for condensation}
In this section we aim to calculate the order parameter  $\langle \mathcal{O}\rangle$ for the $s$-wave condensate in the boundary field theory. In order to do so, we need to consider the behaviour of the gauge field $\phi$ near the critical temperature $T_{c}$. Substituting Eqn.\eqref{e18} into Eqn.\eqref{e10} we may find
\begin{equation}
\phi''(z) +\dfrac{2bz^{3}}{r_{+}^{2}}\phi'^{3}(z)=\dfrac{\mathcal{F}^{2}(z)z^{2}\langle \mathcal{O}\rangle ^{2}}{r_{+}^{4}(1-z^{3})}\left(1-\dfrac{3bz^{4}\phi'^{2}(z)}{2r_{+}^{2}}\right)\phi(z)+{\bf{O}}(b^{2})
\label{eq4.1}
\end{equation}

It is to be noted that in the subsequent analysis only terms upto linear order in $b$ have been considered.

As a next step, we expand $\phi(z)$ perturbatively in the small parameter $\langle \mathcal{O}\rangle^{2}/r_{+}^{4}$ as follows :
\begin{equation}
\frac{\phi(z)}{r_{+}}=\frac{\phi_{0}(z)}{r_{+}}+\dfrac{\langle \mathcal{O}\rangle^{2}}{r_{+}^{4}}\chi(z)+\textit{higher order terms}.
\label{eq4.2}
\end{equation}

where $\phi_{0}$ is the solution of Eqn.\eqref{e13}. Here $\chi(z)$ is some arbitrary function which satisfies the boundary condition
\begin{equation}
\chi(1)=\chi'(1)=0.
\label{eq4.2a}
\end{equation}

Substituting Eqn.\eqref{eq4.2} and Eqn.\eqref{e14} we may write Eqn.\eqref{eq4.1} in terms of $\chi(z)$ as
\begin{equation}
\chi''(z)+6b\lambda^{2}z^{3}\xi'^{2}(z)\chi'(z)=\lambda\dfrac{\mathcal{F}^{2}(z)z^{2}\xi(z)}{(1-z^{3})}\left(1-\dfrac{3b\lambda^{2}z^{4}\xi'^{2}(z)}{2}\right)
\label{eq4.3}
\end{equation}

Multiplying both sides of Eqn.\eqref{eq4.3} by $e^{3b\lambda^{2}z^{4}\xi'^{2}(z)/2}$ and considering terms upto order $b$, we obtain\footnote{The detailed derivation of this equation is given in the Appendix.}
\begin{eqnarray}
\frac{d}{dz}\left(e^{3b\lambda^{2}z^{4}\xi'^{2}(z)/2}\chi'(z)\right)&=&e^{3b\lambda^{2}z^{4}\xi'^{2}(z)/2}\lambda\dfrac{\mathcal{F}^{2}(z)z^{2}\xi(z)}{(1-z^{3})}\left(1-\dfrac{3b\lambda^{2}z^{4}\xi'^{2}(z)}{2}\right)\nonumber\\
&=&\lambda\dfrac{\mathcal{F}^{2}(z)z^{2}\xi(z)}{(1-z^{3})}.
\label{eq4.4}
\end{eqnarray}
Using the boundary condition Eqn.\eqref{eq4.2a} and integrating Eqn.\eqref{eq4.4} in the interval $[0,\,1]$ we finally obtain
\begin{equation}
\chi'(0)=-\lambda \,(\mathcal{A}_{1}+\mathcal{A}_{2})
\label{eq4.5}
\end{equation}
where, 
\begin{eqnarray}
\mathcal{A}_{1}&=&\int_{0}^{\Lambda}\dfrac{\mathcal{F}^{2}(z)z^{2}\xi_{1}(z)}{(1-z^{3})}\;\;\;\;\;\;\;\;\;\;\;,\textit{for}\;\; 0\leq z< \Lambda\nonumber\\
&=&\frac{1}{12600\Lambda^{4}}\lbrace -70\sqrt{3}\pi(-1-10
\Lambda^{4}+\alpha(-2+\Lambda^{4}(10\alpha +18(2+\alpha)\Lambda -10(2+\alpha)\Lambda^{2}+\nonumber\\
&&(2+\alpha)\Lambda^{6})))+\Lambda(126\Lambda(-5+98\Lambda^{3})+
30\alpha(84+\Lambda^{3}(21+2\Lambda^{3}(244+21\Lambda(-10\nonumber\\
&&+\Lambda^{4}))))-35\alpha^{2}\Lambda^{2}(12+\Lambda^{3}(474+\Lambda(-360+\Lambda^{2}(94+9\Lambda (-10+4\Lambda+\Lambda^{4}))))))\nonumber\\
&&-420\;\log(1-\Lambda)+210\;log(1+\Lambda+\Lambda^{2})+210(2\sqrt{3}(-1-10\Lambda^{4}+\alpha(-2+\Lambda^{4}(10\alpha \nonumber\\
&&+18(2+\alpha)\Lambda -10(2+\alpha)\Lambda^{2}+(2+\alpha)\Lambda^{6})))\;tan^{-1}
\left(\frac{1+2\Lambda}{\sqrt{3}}\right)-(-2\alpha -10(1+\alpha^{2})\nonumber\\
&&\Lambda^{4}+18(\alpha -2)\alpha \Lambda^{5}-10(\alpha -2)\alpha \Lambda^{6}+(\alpha -2)\alpha\Lambda^{10})(2\;log(1-\Lambda)-\;\log(1+\Lambda +\Lambda^{2}))\nonumber\\
&&-2(\alpha^{2}+20\alpha \Lambda^{4}+\Lambda^{5}(18-10\Lambda +\Lambda^{5}))\;\log(1-\Lambda^{3}))\rbrace 
\label{eq4.6}
\end{eqnarray}
and
\begin{eqnarray}
\mathcal{A}_{2}&=&\int_{\Lambda}^{1}\dfrac{\mathcal{F}^{2}(z)z^{2}\xi_{2}(z)}{(1-z^{3})}\;\;\;\;\;\;\;\;\;\;\;,\textit{for}\;\; \Lambda< z\leq 1\nonumber\\
&=&\dfrac{\Lambda^{2}}{360}\lbrace 4\sqrt{3}\pi(-10(1+\alpha(4+\alpha))+(-1+2\alpha(1+\alpha))
\Lambda^{4})+12\sqrt{3}(10+10\alpha(4+\alpha)+\Lambda^{4}\nonumber\\
&&-2\alpha(1+\alpha)\Lambda^{4})\;
tan^{-1}\left(\frac{1+2\Lambda}{\sqrt{3}}\right)+3(-12\alpha (-10+\Lambda(20-10\Lambda +\Lambda^{5}+\Lambda^{3}(-1+\;\log 3)))\nonumber\\
&&-6(\Lambda^{2}+\Lambda^{4}(-1+\;\log 3))
+\alpha^{2}(110+\Lambda(-120+\Lambda^{2}(40+3\Lambda(-15+4\Lambda+\Lambda^{4})))-60\;\log 3)\nonumber\\
&&+60\;\log 3-24\alpha\Lambda^{4}
\log\Lambda+6(-10+\Lambda^{4}+2\alpha(5\alpha+\Lambda^{4}))\;\log(1+
\Lambda +\Lambda^{2}))\rbrace 
\label{eq4.7}
\end{eqnarray}
where $\xi_{1}(z)$ and $\xi_{2}(z)$ were identified earlier (Eqn.\eqref{e15} and Eqn.\eqref{e17}).

Now from Eqn.\eqref{e11} and Eqn.\eqref{eq4.2} we may write
\begin{eqnarray}
\frac{\mu}{r_{+}}-\frac{\rho}{r_{+}^{2}}z&=&\frac{\phi_{0}(z)}{r_{+}}+\dfrac{\langle \mathcal{O}\rangle^{2}}{r_{+}^{4}}\chi(z)\nonumber\\
&=&\lambda \;\xi(z)+\dfrac{\langle \mathcal{O}\rangle^{2}}{r_{+}^{4}}\left\lbrace \chi(0)+z\;\chi'(0)+\frac{z^{2}}{2!}\chi''(0)+... \right\rbrace 
\label{eq4.8}
\end{eqnarray}

It is to be noted that, while writing the r.h.s of Eqn.\eqref{eq4.8} we have made a Taylor expansion of $\chi(z)$ around $z=0$.

Comparing the coefficients of $z$ from both sides of Eqn.\eqref{eq4.8}, we obtain
\begin{equation}
\frac{\rho}{r_{+}^{2}}=\lambda -\dfrac{\langle \mathcal{O}\rangle^{2}}{r_{+}^{4}}\chi'(0).
\label{eq4.9}
\end{equation}


Substituting Eqn.\eqref{eq4.5} we may write Eqn.\eqref{eq4.9} in the following form:
\begin{equation}
\frac{\rho}{r_{+}^{2}}=\lambda\left\lbrace 1+\dfrac{\langle \mathcal{O}\rangle^{2}}{r_{+}^{4}}(\mathcal{A}_{1}+\mathcal{A}_{2})\right\rbrace.
\label{eq4.10}
\end{equation}

Substituting $\lambda =\rho /r_{+(c)}^{2}$ (cf. Eqn.\eqref{e14}) into Eqn.\eqref{eq4.10} we finally obtain the expression for the order parameter $\langle\mathcal{O}\rangle$ near the critical temperature ($T_{c}$) as,
\begin{equation}
\langle\mathcal{O}\rangle =\beta\; T_{c}^{2}\sqrt{1-\frac{T}{T_{c}}}
\label{eq4.11}
\end{equation}
where the coefficient $\beta$ is given by,
\begin{equation}
\beta =\dfrac{16\sqrt{2}\;\pi^{2}}{9\sqrt{(\mathcal{A}_{1}+\mathcal{A}_{2})}}.
\label{eq4.12}
\end{equation}

In the following table (Table 2) we have provided both analytic as well as numerical\cite{hs19} values for the coefficient $\beta$ corresponding to different values of the Born-Infeld parameter ($b$). 
\begin{table}[h]
\caption{Values of $\beta$ (Eqn.\eqref{eq4.12}) for different values of $b$}   
\centering                          
\begin{tabular}{c c c c c c}            
\hline\hline                        
Values of $b$ & Values of $\alpha$ & ($\mathcal{A}_{1}+\mathcal{A}_{2}$)  & $\beta_{SL}$ &  $\beta_{numerical}$  \\ [0.05ex]
\hline
0.1 & 0.653219 & 0.0442811 & 117.919 & 207.360  \\ \\
0.2 & 0.656050 & 0.0388491 & 125.893 & 302.760  \\ \\
0.3 & 0.660111 & 0.0352282 & 132.205 & 432.640 \\  [0.05ex]           
\hline                              
\end{tabular}\label{T2}  
\end{table}

Here (from Table 2) one can note that both the values that are obtained through different approaches are in the same order. The difference that is caused is mainly due to the perturbative technique itself where we have dropped higher order terms in the coupling (b). Similar features have also been found earlier\cite{hs11}. However, the trend is unique, i.e. $\beta$ increases as we increase the value of coupling $b$ (see also Fig.(1)). Indeed, it would be interesting to carry out the analysis taking into account higher order terms in the coupling $b$ which is expected to reduce the disparity between the analytic and numerical results.
\begin{figure}[h]
\begin{minipage}[b]{0.5\linewidth}
\centering
\includegraphics[angle=0,width=8cm,keepaspectratio]{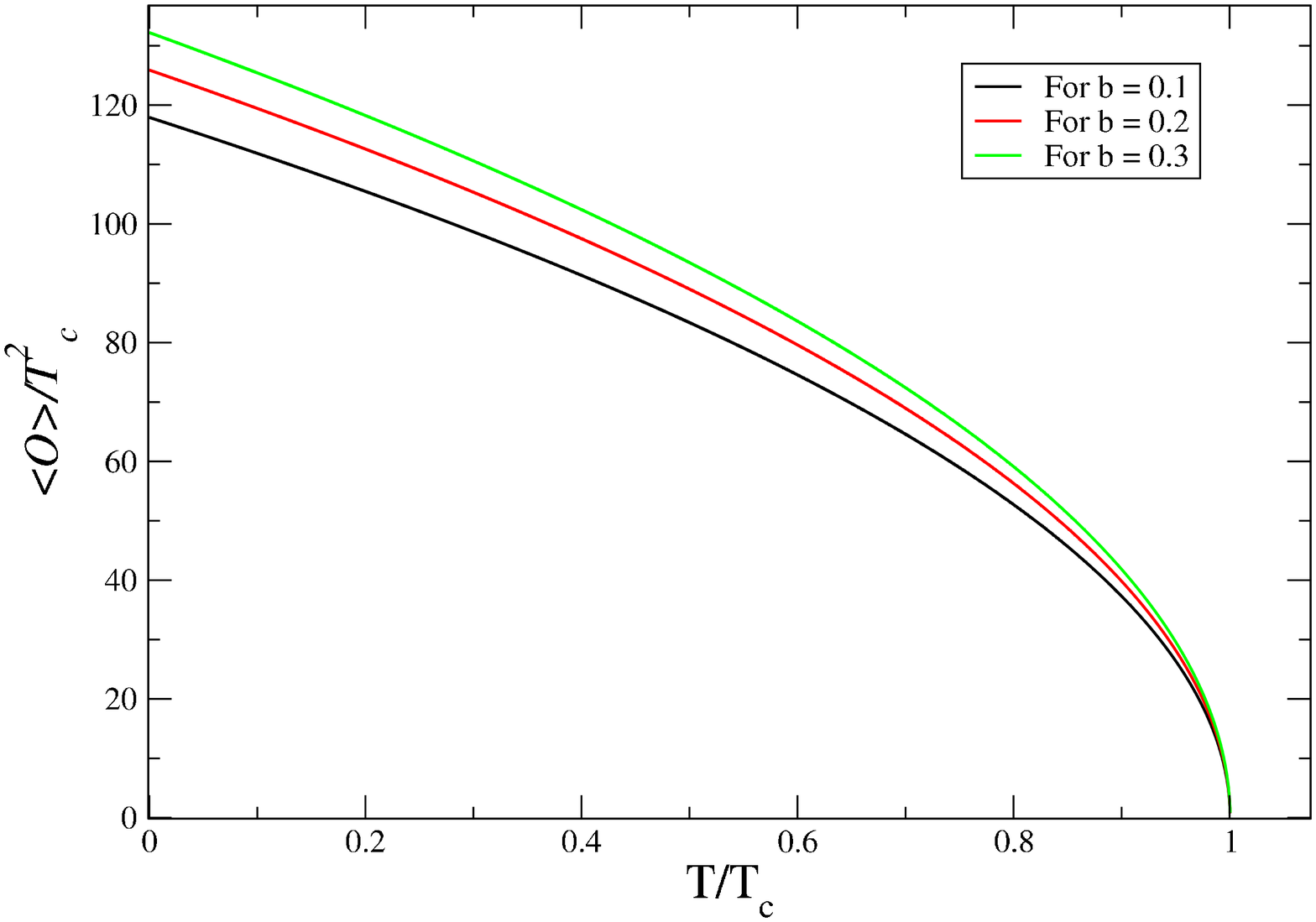}
\caption[]{\it Plot of $\langle\mathcal{O}\rangle /T_{c}^{2}$ with $T/T_{c}$ for different values of $b$.}
\label{figure 1}
\end{minipage}
\hspace{.1cm}
\begin{minipage}[b]{0.5\linewidth}
\centering
\includegraphics[angle=0,width=8cm,keepaspectratio]{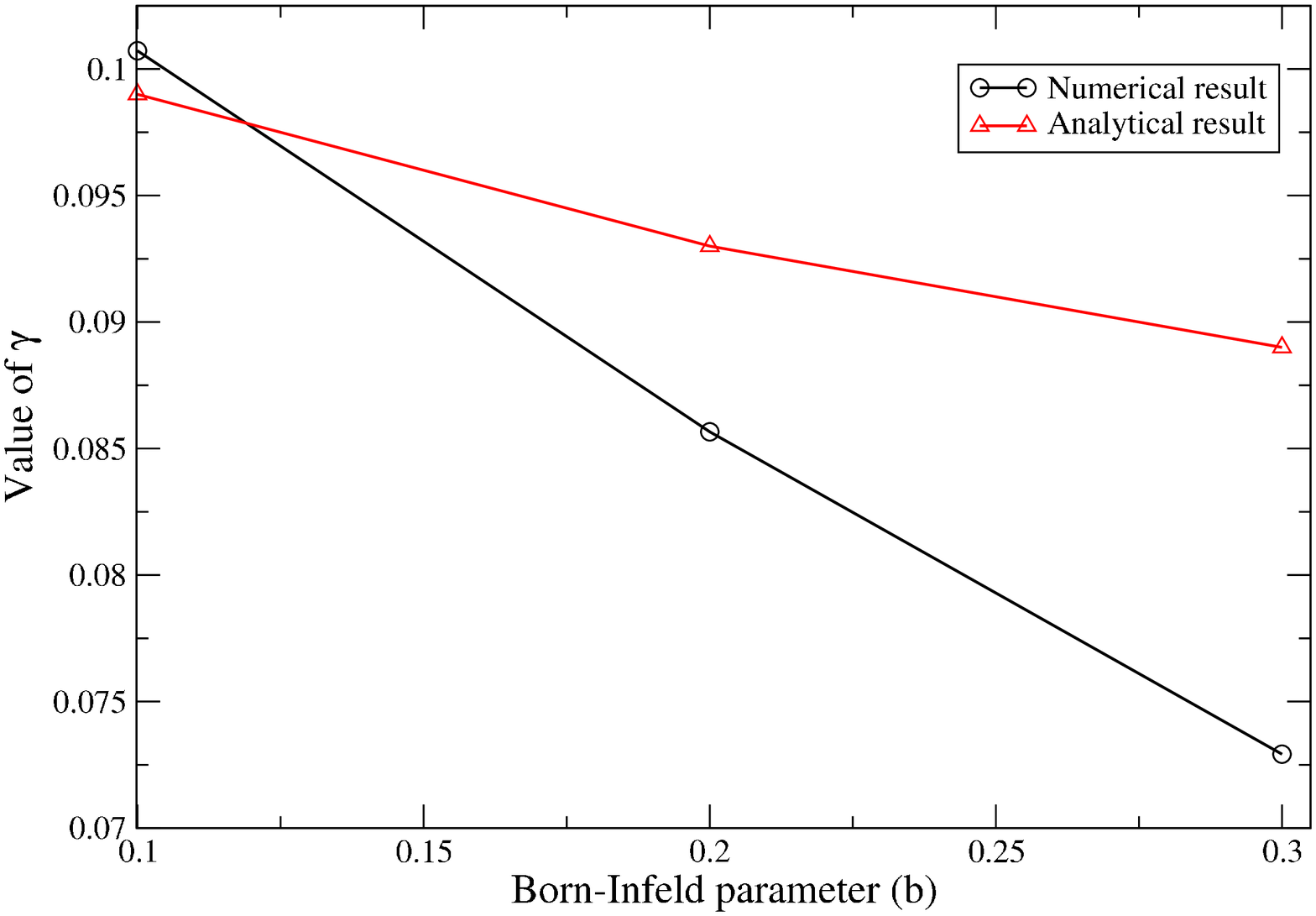}
\caption[]{\it Plot of the coefficient of $T_{c}$ ($\gamma$) with the BI parameter ($b$).}
\label{figure 2}
\end{minipage}
\end{figure}

\section{Conclusions}
In this paper we have considered a holographic model of superconductor based on fundamental principles of AdS/CFT duality. Among several models of holographic superconductor in the AdS black hole background, we have taken into account a model in which nonlinear Born-Infeld Lagrangian is included in the matter action. The main purpose for considering the BI theory is that it corresponds to the higher derivative corrections of the gauge fields in the usual Abelian theory that effectively describes the \textit{low energy} behaviour of the string theory. In this sense it may be considered as the generalized version of the Abelian model. These corrections must have nontrivial influences on the physical properties of the system. 

The aim of the present article is to study the effects of these higher derivative corrections on the holographic $s$-wave condensate analytically. In this paper we have been able to extend the so called Sturm-Liouville (SL) method for this nonlinear model. This method was first introduced in \cite{hs7} in the context of usual Maxwell theory. From our analysis it is indeed evident that extending such a method for the nonlinear model creates difficulties in the analysis. However, we have been able to construct an analytic technique based on this SL method in order to analyse the properties of this holographic superconductor subjected to a nontrivial boundary condition. On top of it, our approach reveals the fact that the solutions of the field equations are highly nontrivial and are not even exactly solvable. The analytic method presented here provides a smooth platform to deal with this difficulty.

The novelty of the present paper is that, we have analytically studied the effects of the BI coupling parameter $b$ on the critical temperature and the condensation operator near the critical point. It is observed that the above physical quantities are indeed affected due to the higher derivative corrections. The results thus obtained from our calculations can be summarized qualitatively as follow:

$\bullet$  The critical temperature ($T_{c}$) increases as we decrease the value of $b$ indicating the onset of a harder condensation (Table 1).

$\bullet$ The value of the order parameter increases with the increase of $b$ (Table 2).

The point that must be stressed at this stage of discussion is that the analytic approach is always more preferable than the numerical approach. This is due to the fact that the numerical results become less reliable when the temperature $T$ approaches to zero\cite{hs3,hs7}. In this temperature limit the numerical solutions to the nonlinear field equations becomes very much difficult and therefore the determination of the nature of the condensate becomes practically very arduous unless analytic methods are taken into account. Therefore, analytic method is always more reliable while performing computations as $T\rightarrow0$.

The deviation of the analytic values from those of the numerical one (Tables 1 and 2) is not unusual\cite{hs11}, considering the difference in the two approaches (analytic and numerical). Contrary to the numerical approach, in the analytic method we have taken into account only the leading order terms in the coupling $b$. Certainly, there is a great amount of approximation involved which is absent in the numerical technique.
 
The difference between the two approaches in fact motivates us to enquire into a more general analytic approach in which the above disparity may be reduced and the agreement eventually becomes more close. Apart from this there are other possibilities which we must emphasize in order to obtain enriched physics from the theoretical point of view. These may be stated as follows:

(i) It would be very much interesting to repeat the above analysis in the presence of back reaction.

(ii) With the mathematical technique presented here one can perform the analysis in higher dimensions.

(iii) Apart from the BI electrodynamics one can also analyse the problem considering any other non-linear theory (theory considering power Maxwell action, Hoffman-Infeld theory, logarithmic electrodynamics etc.) existing in the literature.   

As final remarks, we would like to mention that the numerical results obtained in the existing literature have always been substantiated by analytic results. However, one may confirm the validity of the analytic results obtained by the Sturm-Liouville (SL) method (without referring to the numerical results) by comparing them with the results obtained from an alternative analytic technique which is known as the matching method\cite{hs8}. 

\section*{Appendix}
\begin{flushleft}
\underline{\bf{Derivation of Eqn.\eqref{eq4.4}:}}
\end{flushleft}
The {\bf{l.h.s}} of Eqn.\eqref{eq4.4} may be written as,
\begin{equation}
\frac{d}{dz}\left(e^{3b\lambda^{2}z^{4}\xi'^{2}(z)/2}\chi'(z)\right)=e^{3b\lambda^{2}z^{4}\xi'^{2}(z)/2}\left[\chi''(z)+6b\lambda^{2}z^{3}\xi'^{2}(z)\chi'(z)+3b\lambda^{2}z^{4}\xi'(z)\xi''(z)\chi'(z)\right].
\label{A_1}
\end{equation}

The last term in the r.h.s of Eqn.\eqref{A_1} can be rewritten as,
\begin{eqnarray}
3b\lambda^{2}z^{4}\xi'(z)\xi''(z)\chi'(z)&=&3b\lambda^{2}z^{4}
\xi'(z)\left(\frac{\phi_{0}''(z)}{\lambda r_{+}}\right)\chi'(z)\nonumber\\
&=&\dfrac{-6b^{2}\lambda z^{7}\xi'(z)}{r_{+}^{3}}\phi_{0}'^{3}(z)\chi'(z)\nonumber\\
&=&-6b^{2}\lambda^{4}z^{7}
\xi'^{4}(z)\chi'(z)\nonumber\\
&\approx &0.
\label{A_2}
\end{eqnarray}
where we have used Eqn.\eqref{e13}.

Therefore Eqn.\eqref{A_1} becomes
\begin{equation}
\frac{d}{dz}\left(e^{3b\lambda^{2}z^{4}\xi'^{2}(z)/2}\chi'(z)\right)=e^{3b\lambda^{2}z^{4}\xi'^{2}(z)/2}\left[\chi''(z)+6b\lambda^{2}z^{3}\xi'^{2}(z)\chi'(z)\right].
\label{A_3}
\end{equation}

The {\bf{r.h.s}} of Eqn.\eqref{eq4.4} may be written as, 
\begin{eqnarray}
e^{3b\lambda^{2}z^{4}\xi'^{2}(z)/2}\dfrac{\lambda\mathcal{F}^{2}(z)z^{2}\xi(z)}{(1-z^{3})}\left(1-\dfrac{3b\lambda^{2}z^{4}\xi'^{2}(z)}{2}\right)&\approx&\left(1+\dfrac{3b\lambda^{2}z^{4}\xi'^{2}(z)}{2}\right)\dfrac{\lambda\mathcal{F}^{2}(z)z^{2}\xi(z)}{(1-z^{3})}\nonumber\\
&&\left(1-\dfrac{3b\lambda^{2}z^{4}\xi'^{2}(z)}{2}\right)\nonumber\\
&\approx&\dfrac{\lambda\mathcal{F}^{2}(z)z^{2}\xi(z)}{(1-z^{3})}.
\label{A_4}
\end{eqnarray}

Combining Eqn.\eqref{A_3} and Eqn.\eqref{A_4} we obtain the desired form of Eqn.\eqref{eq4.4}.

\section*{Acknowledgements}
AL and DR would like to thank CSIR, Government of India, for financial support. SG would like to thank IUCAA, Pune, India.

\end{document}